\documentclass[a4paper]{article}
\usepackage[n, advantage, operators,sets, adversary,landau,probability,notions,logic,ff,mm,primitives,events, complexity,asymptotics,keys]{cryptocode}
\usepackage{cryptocode}
\usepackage{algpseudocode}
\usepackage{amsmath}
\usepackage{mathtools}
\usepackage[english]{babel}
\usepackage{hyperref}
\usepackage{pgfplots}
\usepackage{tikz}
\usepackage{url}
\hypersetup{
	linktoc=all
}

\usepackage{color}
\definecolor{blue}{rgb}{0.04, 0.06, 0.66}
\definecolor{green}{rgb}{0.12, 0.58, 0}
\usepackage{listings}
\title{A "Final" Security Bug}
\author{Nguyen Thoi Minh Quan
\footnote{https://www.linkedin.com/in/quan-nguyen-a3209817, https://github.com/cryptosubtlety,
https://scholar.google.com/citations?user=9uUqJ9IAAAAJ, msuntmquan@gmail.com}}

\begin{document}
\date{}
\maketitle
\begin{abstract}
This article discusses a fixed \footnote{The fix is public in github \cite{ed25519githubfix} and the bug has been simplified and transformed into a challenge in Google CTF final \cite{ed25519ctf}.} critical security bug in Google Tink's Ed25519 Java implementation. The bug allows \textit{remote attackers} to extract the private key with only two Ed25519 signatures. The vulnerability comes from the misunderstanding of what \textit{"final"} in Java programming language means. The bug was discovered during security review before Google Tink was officially released. It reinforces the challenge in writing safe cryptographic code and the importance of the security review process even for the code written by professional cryptographers.
\end{abstract}

\section{Ed25519}
In this section, we'll briefly describe Ed25519 signature \cite{ed25519}, \cite{ed25519python}. Ed25519 signature is carefully designed with excellent security engineering consideration. You can read a nice insightful article about elliptic-curve signature design by Daniel Berstein and Tanja Lange \cite{ecsignaturedjb}. We use notation from \cite{ed25519}, \cite{ed25519python}.

Fix a prime $q = 2^{255} - 19$, a finite field $F_q$, $d = -121655/121666$. The elliptic curve $E$ is defined by the following equation: $-x^2 + y^2 = 1 + dx^2y^2$. The base point $B$ has prime order $l = 2^{252} + 27742317777372353535851937790883648493$.

Before describing how to compute Ed25519 signatures, we define a few parameters. Fix the bit length $b = 256$. A point $P = (x, y)$ is encoded as $(b-1)$-bit encoding of $y$ followed by a sign bit of $x$, denoted as $\underline{(x, y)}$. We use a cryptographic hash function H = SHA512 so that H's output is 2b-bit length. The private key is $k$. From $H(k) = (h_0, h_1, \cdots, h_{2b - 1})$, we compute $a = 2^{b - 2} + \sum_{i = 3}^{b - 3}2^ih_i$. The public key is $\underline{A}$ where $A = aB$. 

To compute the signature of message $M$, we compute $r = H(h_b, \cdots, h_{2b - 1}, M)$, $R = rB$ and $S = (r + H(\underline{R}, \underline{A}, M) a) \mod l$. The signature is 2b-string $(\underline{R}, \underline{S})$.

\section{The "final" vulnerability}
\subsection*{Security review} Let's talk about the security review process before digging into the vulnerability itself. Google Tink's Java Ed25519 was implemented from scratch because Java Cryptography Architecture (JCA) didn't have Ed25519. The code was complicated, so I conducted a security review for it. It was my 1st time to review optimized low-level cryptographic code, so it was chaotic. The followings did not happen in chronological order as I tried to make the review process look systematic :) 

Believe it or not, whenever I review cryptographic code, the 1st thing I do isn't looking for cryptographic bugs. The reason is that general security bugs such as memory corruption bugs are easier to exploit and have far reaching serious security consequences. If attackers get remote code execution in our process then all bets are off, the cryptographic code's security is not important anymore. Fortunately, in this case, the code was written in memory safe programming language Java, so I could focus on cryptographic bugs.

To me, the optimized code is not intuitive nor understandable. In a typical security review, I read the original papers to understand the protocol and read the code and try to map the code to the papers. For optimized cryptographic code, reading the original papers is not enough, I need to read other papers that describe fast and optimized algorithms. Ed25519's implementation needs optimization at 2 layers : at the finite field layer $F_q$ and at the elliptic curve arithmetic computation layer on $E$. The optimized algorithms make the code incomprehensible. In around 2 months, I spent around 1-2 hours every day. I read the code, read the papers, tried to reason every single line of code what it means. I was pretty clueless what types of security bugs I was looking for. In a typical case, I often read the existing vulnerabilities and attacks to make sure that the code doesn't make the same mistakes. In this case, at the time, there was no precedent for Ed25519 security bugs that I was aware of, i.e., I had nothing to learn from. However, my instinct told me that a complicated code that has been developed in a short period of time and the code was too difficult to understand, it couldn't be right. Therefore, I kept looking and in the end, I had a pretty good idea what's going on with the code at the high level.

I've learned from other security reviews that arithmetic errors in cryptographic code may be exploitable, so I paid attention to the correctness of the code. Besides checking the code, I wrote tests. At the finite field layer $F_q$, I wrote simple tests, yet powerful to give me certain confidence on the correctness of the code. As everyone in security knows, Java BigInteger is not safe to use in cryptographic implementation but I could use its arithmetic computation results to test against Tink's finite field implementation. If BigInteger computation's results match Tink $F_q$ computation's results, I can sleep a little bit better. I didn't find any bugs in Tink $F_q$'s implementation. At the elliptic curve arithmetic computation layer, I found an apparently non-exploitable bug. This is the bug's description in github "The bug is that isNonZeroVarTime assumes the input is reduced while it isn't. Furthermore, the reduced number representation is not unique, e.g., the following array is essentially zero [67108845, 33554431, 67108863, 33554431, 67108863, 33554431, 67108863, 33554431, 67108863, 33554431]. We have to call Field25519.contract() before checking for zero." I couldn't recall how I found those mysterious numbers, probably through a tedious process of printing out certain variables or solving some simple equations. Besides testing the code with Google Wycheproof's tests, I also wrote a few other tests. In one test, I fixed a key, generated random multiple messages, signed them and verified their signatures. This is to test 2 properties: one key can be used to sign multiple messages and sign and verify functions are compatible with each other. In another test, I used one key to sign the same message multiple times and as Ed25519 is deterministic, the outputs should be a single unique signature. This is to test the deterministic property of Ed25519 signature. One of my tests failed and during the investigation of the failed test, I found the "final" bug which I will describe in the next section. It's worth mentioning that standard tests using the test vectors in RFC won't detect the bug as I'll explain below. In fact, the code's author wrote tests using RFC test vectors.

\subsection*{The "final" bug}

Besides its critical severity, the bug is fascinating because it's unexpected and it happened at a place where I typically never looked at. It reminds me of fond memory hunting for security bugs.

Let's look at the extracted vulnerable code.

\begin{lstlisting}[language=Java]
public byte[] sign(final byte[] data) {
    return Ed25519.sign(data, publicKey, hashedPrivateKey);
}

static byte[] sign(final byte[] message, final byte[] publicKey,
        final byte[] hashedPrivateKey) {
    MessageDigest digest = EngineFactory.MESSAGE_DIGEST
        .getInstance("SHA-512");
    digest.update(hashedPrivateKey, FIELD_LEN, FIELD_LEN);
    digest.update(message);
    byte[] r = digest.digest();
    reduce(r);
    byte[] rB = Arrays.copyOfRange(scalarMult(r).toBytes(), 0,
            FIELD_LEN);
    digest.reset();
    digest.update(rB);
    digest.update(publicKey);
    digest.update(message);
    byte[] hram = digest.digest();
    reduce(hram);
    mulAdd(hashedPrivateKey, hram, hashedPrivateKey, r);
    return Bytes.concat(rB, Arrays.copyOfRange(hashedPrivateKey, 0,
                FIELD_LEN));
}

// Computes (ab + c) mod l.
// Note that the method only uses the 1st 32 bytes of each array. 
private static void mulAdd(byte[] s, byte[] a, byte[] b, byte[] c)

\end{lstlisting}

Recall that we need to compute $(h_0, h_1, \cdots, h_{2b - 1}) = H(k)$ where $k$ is the private key. In signature computation, the private key $k$ is never used directly, so the code's author precomputes $(h_0, h_1, \cdots, h_{2b - 1})$ once and stores the result in the variable \emph{hashedPrivateKey}. The 1st \emph{sign(byte[] data)} method calls the 2nd \emph{sign(final byte[] message, final byte[] publicKey, final byte[] hashedPrivateKey)} method. Pay attention to the keyword \emph{"final"} in method's parameter \emph{final byte[] hashedPrivateKey}. In Java programming language, the keyword \emph{final}  signals the intention that the parameter is a constant and the callee shouldn't change the parameter. However, while the declaration \emph{final byte[] x} means that \emph{x} is constant, as $x$ is just a reference to a "byte[]" array, \emph{x}'s content (aka the array) can be changed! I.e., if the callee changes elements of the array \emph{hashedPrivateKey}, the compiler will not raise compiling errors. 

Let's continue exploring the code. The 2nd \emph{sign} method first computes $H(R, publicKey, message)$ and stores the result in variable \emph{hram}. It then calls \emph{mulAdd(hashedPrivateKey, hram, hashedPrivateKey, r)}. The \emph{mulAdd(s, a, b, c)} method computes $(a*b + c) \mod l$, i.e., it computes $(hram*hashedPrivateKey[:32] + r) \mod l$ (where $hashedPrivateKey[:32]$ denotes the 1st 32 bytes of $hashedPrivatekey$), i.e., it's our $S$. There is an odd thing happening here as well. The variable \emph{hashedPrivateKey} is reused twice in the method \emph{mulAdd(hashedPrivateKey, hram, hashedPrivateKey, r)}, i.e., after computation, it stores the result back to \emph{hashedPrivateKey}, i.e., it stores $S$ backs to the first 32 bytes of \emph{hashedPrivateKey} (recall that \emph{hashedPrivateKey} is an 64-byte array). In combination of the \emph{final} issue above, we see that:

\begin{itemize}
\item \emph{hashedPrivateKey} value is changed.
\item The 1st 32 bytes of \emph{modified} hashedPrivateKey is published in $S$. Note that the 1st 32 bytes of \emph{original} hashedPrivateKey is not leaked.
\end{itemize}

Once the bug is found, the exploitation is pretty straightforward. In fact, based on the bug, I wrote a challenge in Google CTF final \cite{ed25519ctf} and several CTF teams could solve it. The challenge was named Ed25519Final where the word "final" has double meanings: the challenge was used in the CTF \emph{final} and it gave a hint where the bug was, in the "final" keyword in Java programming language.

Now, let's describe how to compute $a$ if we use the same key to sign the same message \emph{twice}. We have the following equations:
\begin{align*}
S_1 = (r_1 + H(R, publicKey, message) * a_1) \mod l \\
S_2 = (r_2 + H(R, publicKey, message) * a_2) \mod l  \\
\end{align*}
where $a = a_1, a_2$ correspond to the 1st 32 bytes of our \emph{hashedPrivateKey}. Before continuing, it's worth noting that the 1st signature is correctly produced and only the 2nd signature is wrong, so the standard tests in RFC won't help detecting the bug. We notice the following:
\begin{itemize}
\item As we sign the same message: $r_1 = r_2 = r$.
\item Due to the leakage and modification of $hashedPrivateKey$, we know that $a_2 = S_1$.
\end{itemize}
Therefore:
\begin{align*}
S_1 = (r + H(R, publicKey, message) * a_1) \mod l \mbox{ (1)} \\
S_2 = (r + H(R, publicKey, message) * S_1) \mod l \mbox{ (2)} \\
\end{align*}
From equation (2), we can compute $r = (S_2 - H(R, publicKey, message) * S_1) \mod l $. Plug $r$ into (1), we have $a = a_1 = ((S_1 - r) * (H(R, publicKey, message)^{-1} \mod l)) \mod l$. Happy hacking!
\section*{Acknowledgement}
Thanks Daniel Bleichenbacher and Thai Duong for fruitful cryptographic discussions during Google Tink and Wycheproof's development.
\bibliographystyle{unsrt}
\bibliography{research}
\end{document}